# Cooperative Dispatch of Microgrids Community Using Risk-Sensitive Reinforcement Learning with Monotonously Improved Performance

Ziqing Zhu, *Member, IEEE*, Xiang Gao, Siqi Bu, *Senior Member, IEEE*, Ka Wing Chan, *Member, IEEE*, Bin Zhou, *Senior Member, IEEE*, Shiwei Xia, *Senior Member, IEEE*

*Abstract*—The integration of individual microgrids (MGs) into Microgrid Clusters (MGCs) significantly improves the reliability and flexibility of energy supply, through resource sharing and ensuring backup during outages. The dispatch of MGCs is the key challenge to be tackled to ensure their secure and economic operation. Currently, there is a lack of optimization method that can achieve a trade-off among top-priority requirements of MGCs' dispatch, including fast computation speed, optimality, multiple objectives, and risk mitigation against uncertainty. In this paper, a novel Multi-Objective, Risk-Sensitive, and Online Trust Region Policy Optimization (RS-TRPO) Algorithm is proposed to tackle this problem. First, a dispatch paradigm for autonomous MGs in the MGC is proposed, enabling them sequentially implement their self-dispatch to mitigate potential conflicts. This dispatch paradigm is then formulated as a Markov Game model, which is finally solved by the RS-TRPO algorithm. This online algorithm enables MGs to spontaneously search for the Pareto Frontier considering multiple objectives and risk mitigation. The outstanding computational performance of this algorithm is demonstrated in comparison with mathematical programming methods and heuristic algorithms in a modified IEEE 30-Bus Test System integrated with 4 autonomous MGs.

*Index Terms*—Microgrid clusters, distributed dispatch, multi-agent reinforcement learning, multi-objective optimization, optimization under uncertainty

## I. INTRODUCTION

### A. General Background

**M**icrogrids (MGs) are groups of interconnected energy resources and loads that are linked with the local active distribution network (ADN) or main grid, being capable of operating independently and providing energy to specific area [1]. Particularly, MGs allow for greater energy independence and resilience [2], as it can be used to provide energy during times of outages or natural disasters. In addition, MGs can help reduce energy costs and emissions, as they are often powered by renewable distributed energy resources (RDERs) such as photovoltaics (PVs) and wind turbines [3]. Due to these superiorities, the clustering of individual MGs into MGCs (MGCs) is an innovative way for maximizing synergistic effects when MGs are located in close proximity to one another, because such an aggregation can increase the reliability and flexibility of energy supply by allowing the sharing of energy resources among MGs, and thus providing a backup in the event of power outages [4].

### B. Literature Review of MGCs' Dispatch

The coordinated dispatch (defined as "energy management" in some literatures) of multiple MGs is the main challenge of MGCs' operation, with the following open questions to be tackled. **First**, the uncertainty of RDERs' generation output brings huge difficulties to the real-time power balancing. This is an old topic discussed by considerable number of literatures, with solutions including the stochastic and robust optimization. A distributed robust model predictive control (DRMPC)-based strategy is proposed in [5] for real-time energy management of MGCs, by combining the advantages of robust optimization and model predictive control. In [6], a cumulative relative regret decision-making strategy is proposed to hedge uncertainties in MGC operation with consideration of the delay of heat transfer and the fuzziness of heating comfort. A two-stage stochastic programming model is developed in [7] to achieve the interactive energy management between MGCs and the main grid. **Second**, the consideration of multiple objectives, including the improvement of economic efficiency, power quality, reliability and resilience, is necessary but difficult to implement; for example, the trade-off between operation cost and system reliability [8]; the simultaneous consideration of economic efficiency and net-zero [9]; and utilities of different demand-side users [10]. Some heuristic algorithms such as particle swarm algorithm [11], grey wolf algorithm [12] and teaching-learning-based algorithm [13] are deployed to mitigate complicated convexification of multiple objective functions. **Third**, individual MGs are autonomous and should be decentralized dispatched. Therefore, the coordination of these MGs' energy management to mitigate potential conflicts, and energy exchange/trading mechanisms among MGs, are significant topics for investigation. A mixed-integral linear programming (MILP)-based energy management is designated to coordinate multiple MGs in a hierarchical manner [14], considering the trade-off between the local and global optimization. In [15], a transactive energy framework is proposed to facilitate the energy trading among the upstream distribution system operator (DSO) and downstream MGs. Besides, several game-theoretic-based frameworks are proposed in [16] – [17] with consideration of cooperation or competition relationships among MGs. Based on these frameworks, some local energy markets with detailed pricing schemes are established in [18] – [20], to incentivize MGs' energy sharing in the most economically efficient way with maximized social welfare.

However, the aforementioned works still have substantial deficiencies to be addressed. **First**, there lacks method that can handle all these concerns simultaneously. For instance, the risk mitigation against uncertainties relies on stochastic/robust programming which necessitates convexification of objective functions and constraints; however, the consideration of multi-objectives will render the convexification intractable. The solution of multi-objective optimization, i.e., heuristic algorithms, cannot guarantee the effectiveness of risk mitigation. Furthermore, both stochastic/robust programming and heuristic algorithms are centralized optimization methods, and therefore cannot be accommodated into distributed MGC paradigms as proposed in [15] – [20]. **Second**, most of these methods are off-line optimization and cannot respond in a fast speed while the *state* (operation status) is changing (for example, variations of load demand, and unexpected outages).

### C. Potentiality of RL for MGCs' Dispatch

In recent years, the cutting-edge Reinforcement Learning (RL) has emerged as an effective solution to optimization problems. RL is a learning framework that simulates how a group of "smart agents" interact with each other and optimize their own "policies" to receive more profits under specified rules (i.e., the environment) [21]. Users can customize the agents' characteristics (i.e., what decisions to make, how to deal with risks, etc.) and the environment (i.e., how the profit allocated to each agent is determined based on their decisions), to fulfill the users' own simulation requirements. RL has numerous advantages over traditional techniques [22]. **First**, the framework of RL fits perfectly with the distributed operation of MGCs. MGs can be considered as "smart agents", who collaboratively optimize their "policies" of dispatch, to maximize their total profits. **Second**, RL merits faster computation speed because it does not require convexification, which results in heavy computation burdens; instead, RL is a trial-and-error-based optimization method that seeks the optimal policy function in the feasible region through exploration and exploitation, with the goal of maximizing the expected "reward", which is somewhat equivalent to the objective function in conventional optimization problems. Due to this trait, RL is also compatible with multi-objective optimization and risk mitigation, which is accomplished by changing the "reward" function with numerous objectives and penalties for potential risks. **Third**, RL can be deployed to train an online optimizer that can implement the automatic dispatch of MGCs with very little respond time, because RL trains an "optimal policy" that can provide guidance of picking the "optimal solution" at any given state. MARL algorithms not only facilitate the search of global optimality, but also accommodate multi-objective optimization and risk mitigation same as single-agent RL.

Unfortunately, although RL merits these superiorities, **there is still a lack of successful applications of RL in the energy management of MGCs.** For example, the multi-agent version of deep Q-network (MADQN) [23], hierarchical Q-learning (MAHQL) [24], and deep deterministic policy gradient (MADDPG) [25], have been deployed as some initial attempts, but these algorithms still have significant drawbacks. **First**, these attempts have not incorporated multiple objectives as well as risk mitigation which are necessary for MGC operation. **Second**, these algorithms cannot guarantee the monotonous improvement of the policy's performance, and therefore will lead to the sub-optimal solution [26]. **Third**, the simultaneous self-dispatch of multiple MGs may lead to conflicts of power flow and potential violations of constraints, which must be addressed by the DSO with considerable re-dispatch costs [26]. **Fourth**, although RL is similar to optimization problems with decision variables (actions) and objective functions (rewards), the implicit constraints of actions in RL cannot be guaranteed; for example, the nodal voltage constraints in which the power flow must be firstly calculated to obtain the decision variables.

### D. Main Contributions and Paper Structure

This paper aims to fill these aforementioned research gaps by proposing a Risk-Sensitive Trust Region Policy Optimization Algorithm (RS-TRPO) with monotonously improved performance, to achieve the efficient and safe dispatch of MGCs. To the best of authors' knowledge, this is the first work that proposes an online optimization tool to address MGCs' dispatch in a **distributed and collaborative** manner with consideration of **monotonously improved performance guarantee** and **risk mitigation.** Main contributions of this paper are outlined as follows.

- *Dispatch Paradigm:* A distributed-but-cooperative (DBC) dispatch paradigm for autonomous MGs in the MGC is proposed, enabling them sequentially implement their self-dispatch to mitigate potential conflicts.
- *Dispatch Model:* Such a dispatch paradigm is formulated as a Markov game. In this model, autonomous MGs' decision making process are clearly defined, with consideration of MGs' multi-objective optimization and risks due to uncertainties of net load.
- *Dispatch Strategy Optimization:* The RS-TRPO algorithm is proposed to estimate the optimal solution, i.e., the Pareto-Optimal Equilibrium (POE) of the Markov game. The implementation of this algorithm is consistent with the sequential dispatch paradigm, enabling MGs to spontaneously search for the POE with Conditional Value-at-Risk (CVaR)-based risk mitigation.

The remainder of this paper is organized as follows. The distributed and collaborative dispatch paradigm is formulated as Markov Game in Section II. The RS-TRPO algorithm is proposed in Section III. Numerical tests are performed in Section IV. Conclusions are made in Section V.

## II. MODEL FORMULATION

In this section, a DBC dispatch paradigm is proposed to achieve the independent dispatch of autonomous MGs, to achieve the maximization of the total profit in the MGC. The dispatch strategy optimization procedure for each MG is then formulated as a Markov Game model, and the optimality of this model is defined as the Pareto-Optimal Equilibrium (POE), which is the ultimate goal of the RS-TRPO algorithm.

### A. General Description of Dispatch Paradigm

In this paper, we focus on the hourly cooperative dispatch of distributed MGs in the MGC of a distribution network. It is assumed that MGs are invested, maintained and operated by autonomous stakeholders with different purposes, and they implement the self-dispatch rather than controlled by the DSO. However, MGs are assumed to share a common objective to maximize the total profit of the community by satisfying the load demand in the whole community while maintaining the safe operation (i.e., to satisfy the active power balance and

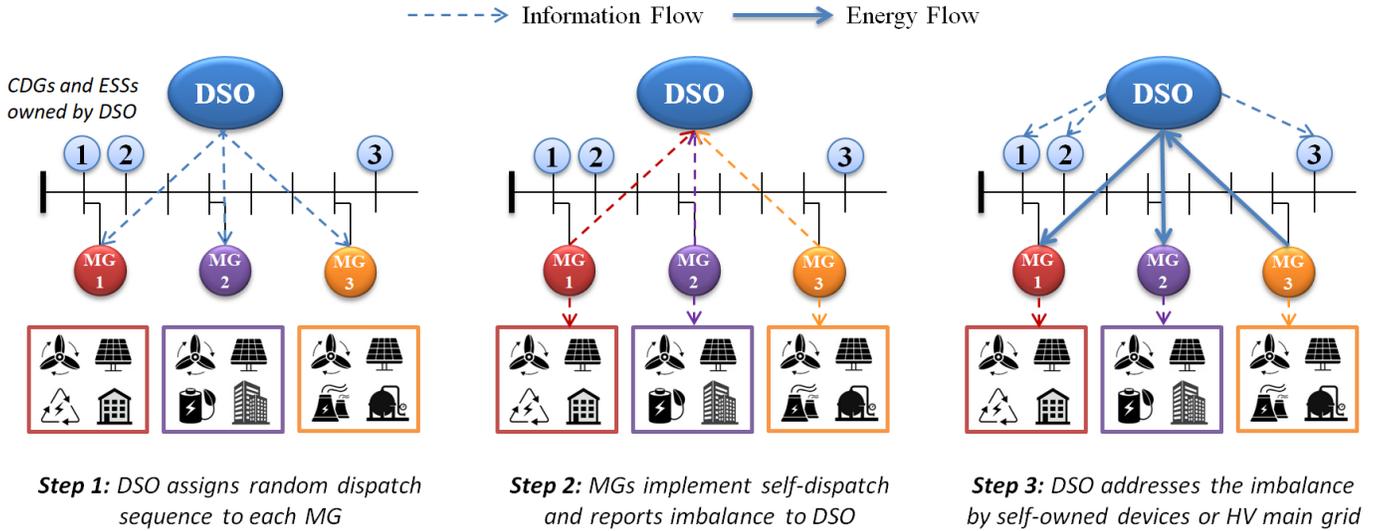

**Fig.1 The Proposed DBC Dispatch Paradigm**

*Step 1:* DSO assigns random dispatch sequence to each MG

*Step 2:* MGs implement self-dispatch and reports imbalance to DSO

*Step 3:* DSO addresses the imbalance by self-owned devices or HV main grid voltage profile limitation); otherwise, the resulted balancing cost or reactive power support from the DSO or the main grid will be allocated to each MG and lead to significant deduction to their own profit.

To achieve the DBC dispatch, a novel sequential dispatch paradigm is proposed as shown in Fig.1. In short, each MG will be allocated their sequence to implement the self-dispatch at each hour, in order to mitigate potential security issues due to power flow conflict if they simultaneously implement the self-dispatch. More detailed descriptions of roles of stakeholders in this paradigm are provided in Table I.

TABLE I
PARTICIPANTS OF MGC IN DISTRIBUTION NETWORKS

| Element | Description |
|---|---|
| *Microgrids (MGs)* | • MGs contain RDERs (including PV and wind turbine), controllable DERs (CDERs, including Gas-turbine, Fuel cell, Combined Heat and Power), as well as load demands.<br>• The RDERs and loads will lead to real-time imbalance and voltage violation, due to the uncertainty of active/ reactive power output.<br>• MGs are required to corporate to mitigate the impact of uncertain net load on the network balancing, which will be addressed by DSO or main grid with expensive fees for deploying balancing services. |
| *Distribution System Operator (DSO)* | • The DSO is responsible for the stable operation of both the distribution network and the downstream MGCs.<br>• In other words, DSO will act after the MGCs' self-dispatch, and then implement the dispatch of controllable generators affiliated to the DSO, to secure the balancing of the whole distribution network and MGCs with minimized cost. |

## B. Markov Game Formulation

Based on the proposed DBC dispatch paradigm, the next challenging question is to model how each MG optimizes their individual dispatch strategy. In this subsection, the Markov Game model is deployed to formulate MGs' dispatch strategy optimization procedure, and the rationales are explained as follows. First, in the hourly dispatch, each MG's decision at hour $t$ will be affected by its decision at $t-1$ due to the generation capacity and ramping capacity limitation. Second, the objective of each MG agent is to maximize the so-called "long-term" revenue, i.e., the total profit in a single day with 24 hourly decisions to be made. This is similar to the dynamic programming (DP), but DP cannot capture the collaborative interactions among different MGs to maximize the total profit in the whole MGC. The Markov Game model merits the advantages of capturing the Markov property, dynamic programming, and interactions among different MGs, and therefore being deployed to formulate the dispatch problem.

The Markov Game model is an agent-based formulation, considering MGs as different smart agents, and modeling their own "dispatch strategy optimization procedure". The model includes the definition of states (i.e., observations of MG agents at each time-step), actions (i.e., decisions of MG agents at each time-step), and rewards (i.e., profits of MG agents at each time-step). The objective of MG agent is to optimize its policy, i.e., action at a given state, with a maximized reward. Table II provides the detailed definition of the above elements, ensuring the consistence between the modeling and the proposed dispatch paradigm in Section II. A.

TABLE II
ELEMENTS OF MARKOV GAME MODEL

| Terminology | Definition | |
|---|---|---|
| | Upstream DSO (Environment) | Downstream MGs (Agents) |
| *Time Step (t)* | Each time step represents one round of hourly dispatch exercise. | |

| Episode (D) | An episode consists with 24 time steps, indicating dispatch exercises in a whole day. |
|---|---|
| State (S) | 1) Total load demand in the distribution network and in each local MG; 2) Available capacity of affiliated CDERs; 3) Forecasted output of affiliated RDERs; 4) Total balancing cost in the previous round of dispatch exercise. |
| Action (A) | The amount of generation of each affiliated CDERs in the distribution network. / The amount of generation of each affiliated CDERs in the local MG. |
| Reward (R) | The total remuneration of satisfying the load demand in the whole distribution network, with consideration of generation costs and balancing costs. |
| Transition ($\mathfrak{T}$) | The probability of next state by taking action $\boldsymbol{a}_t^k$ at the current state. In this paper, we assume that the transition is unknown. |

Based on the formulated Markov game model, each agent will keep optimizing its own policy, and will ideally reach to the convergence with optimality, i.e., the maximization of the total reward in the MGC. Such optimality is herein formally defined as the Pareto-Optimal Equilibrium.

**Definition 1.** *(Pareto-Optimal Equilibrium)* Let $\pi_k := \boldsymbol{S}_k \times \boldsymbol{A}_k \to [0,1]$ denote the policy of the $k$th MG agent, which indicates the probability of taking action $\boldsymbol{A}_k$ at state $\boldsymbol{S}_k$. Let $\boldsymbol{\pi} := \boldsymbol{S}_k \times \boldsymbol{A}_k \times \boldsymbol{A}_{-k} \to [0,1]$ denote the joint policy of all the agents, where $\boldsymbol{A}_{-k}$ represents the actions of agents except for agent $k$. We define the joint policy $\boldsymbol{\pi}$ is the Pareto-Optimal Equilibrium in the Markov game if

$$J_k(\boldsymbol{\pi}|s_k) \geq J_k(\boldsymbol{\pi}_\vartheta|s_k), \forall k, \forall s_k \in \boldsymbol{S}_k, \forall \boldsymbol{\pi}_\vartheta \in \boldsymbol{\Theta} \quad (1)$$

where $\boldsymbol{\pi}_\vartheta$ denotes the arbitrary subset of the available policy set $\boldsymbol{\Theta}$ excluding $\boldsymbol{\pi}$, and $J_k$ is the objective performance of the Markov game, which can be computed as:

$$J_k(\boldsymbol{\pi}|s_k) = E_{\boldsymbol{\pi}}\left\{\sum_{t=0}^{T}\sum_{k=1}^{P}\gamma_k^t \hat{R}_{t+1}(a_k^t \sim \boldsymbol{\pi}|s_k^t)\right\}, \forall k, \forall s_k \in \boldsymbol{S}_k \quad (2)$$

To be more concrete, (2) indicates that by following the joint policy $\boldsymbol{\pi}$, the community can reach the Pareto-Optimal with maximized expected cumulative reward $J_k(\boldsymbol{\pi}|s_k)$, i.e., the total remuneration of satisfying the total load demand in the whole community, and therefore has no incentive to change the joint policy. The common reward $\hat{R}_t$ of MG agents in the community is defined as:

$$\hat{R}_t = \sum_{t\in T}\sum_{k\in K}C_{MG,t}^{load}\hat{P}_{k,t}^{MG} - \sum_{t\in T}\sum_{k\in K}\hat{C}_{MG,t}^{ADN}\hat{P}_{MG,k,t}^{ADN} - \sum_{t\in T}\sum_{k\in K}C_{k,t}^{MG}\hat{P}_{k,t}^{MG} - \sum_{t\in T}C_{curt,t}^{MG}\hat{P}_{curt,t}^{MG} \quad (3)$$

where $C_{MG,t}^{load}$ denotes the price of MGs selling electricity to end users, $\hat{P}_{k,t}^{MG}$ is the total electricity provided by MGs, $\hat{C}_{MG,t}^{ADN}$ denotes the price of DSO providing real-time balancing, $\hat{P}_{MG,k,t}^{ADN}$ is the total electricity provided by the DSO; $C_{k,t}^{MG}$ represents the generation cost of MGs, $C_{curt,t}^{MG}$ and $\hat{P}_{curt,t}^{MG}$ denote the penalty and amount of load curtailment due to the lack of balancing service, respectively.

The set of constraints of actions is defined as:

$$\Gamma_{k,n} := \begin{cases} \Gamma_{k,1}: P_{DG\_min} \leq P_{DG,q,t} \leq P_{DG\_max}, \forall q, \forall t \\ \Gamma_{k,2}: Q_{DG\_min} \leq Q_{DG,q,t} \leq Q_{DG\_max}, \forall q, \forall t \\ \Gamma_{k,3}: P_{ra\_min} \leq P_{DG,q,t} - P_{DG,q,t-1} \leq P_{ra\_max}, \forall q, \forall t \\ \Gamma_{k,4}: \hat{P}_{k,t}^{MG} = \sum_{q\in\mathbb{Q}}P_{DG,q,t} + \sum_{l\in\mathbb{L}}\hat{P}_{RDG,l,t}, \forall t \\ \Gamma_{k,5}: \hat{P}_{curt,t}^{MG} = \sum_{k\in K}\hat{P}_{k,t}^{MG} + \hat{P}_{MG,k,t}^{ADN} - \hat{P}_{MG,t}^{load}, \forall t \end{cases} \quad (4)$$

where $\Gamma_{k,1}$ - $\Gamma_{k,3}$ denote the active power output limitation, reactive power output limitation, and ramping limitation of each CDG, respectively. $\Gamma_{k,4}$ represents that the power output of each MG is the sum of its affiliated CDGs' and RDGs' power output. $\Gamma_{k,5}$ represents that the load curtailment is computed by the difference between total power supply by both MGs and the DSO, and the total load demand.

Note that the terms $\hat{P}_{MG,k,t}^{ADN}$ in (3) are determined by the DSO's dispatching in (3) to secure the network balancing. This is considered as the "environment" of the Markov game, which is defined as follows:

**Definition 2.** *(Environment of the Markov Game)* The environment, i.e., the DSO's dispatch, determines how much reward can be obtained by MG agents after making their decisions (actions). Such a dispatch manner simply follows the rule of merit-order, formulated as follows:

$$\text{Max} \sum_{t\in T}\sum_{i\in I}C_{load,t}^{ADN}P_{DG,i,t} - \sum_{t\in T}\sum_{i\in I}C_{DG,i,t}P_{DG,i,t}$$
$$- \sum_{t\in T}C_{loss,t}^{ADN}L_{t,loss}^{ADN} - \sum_{t\in T}C_{ex,t}^{HV}P_{ex,t}^{HV} - \sum_{t\in T}\left\{\sum_{i\in I}C_{ESS,i,t}P_{ESS,i,t}\right\}$$
$$- \sum_{t\in T}C_{v,t}^{ADN}\sqrt{\alpha\sum_{i\in I}(1-\hat{V}_{i,t})^2 + \beta(1-\hat{V}_{PCC,t})^2} \quad (5)$$

subject to $\Gamma_n :=$

$$\begin{cases} \Gamma_1: V_{min} \leq \hat{V}_{i,t} \leq V_{max}, \forall i, \forall t \\ \Gamma_2: P_{DG\_min} \leq P_{DG,i,t} \leq P_{DG\_max}, \forall i, \forall t \\ \Gamma_3: Q_{DG\_min} \leq Q_{DG,i,t} \leq Q_{DG\_max}, \forall i, \forall t \\ \Gamma_4: P_{ra\_min} \leq P_{DG,i,t} - P_{DG,i,t-1} \leq P_{ra\_max}, \forall i, \forall t \\ \Gamma_5: P_{ESS\_min} \leq P_{ESS,i,t} \leq P_{ESS\_max}, \forall i, \forall t \\ \Gamma_6: SOC_{min} \leq SOC_{ESS,i,t} \leq SOC_{max}, \forall i, \forall t \\ \Gamma_7: \sum_{i\in I}(P_{DG,i,t} + P_{ESS,i,t} + \hat{P}_{RDG,d,t}) + P_{ex,t}^{HV} = P_{load,t}^{ADN} + \sum_{k\in K}P_{MG,k,t}^{ADN} \end{cases}$$

For the objective function, $C_{load,t}^{ADN}$ denotes the price of DSO selling electricity to end users, $C_{DG,i,t}$ and $P_{DG,i,t}$ represent the generation cost and power output of CDGs owned by the DSO, $C_{loss,t}^{ADN}$ and $L_{t,loss}^{ADN}$ denote the reactive power loss and its cost, $C_{ex,t}^{HV}$ and $P_{ex,t}^{HV}$ denote the cost and purchased electricity of real-time balancing service provided by the main grid, $C_{ESS,i,t}$ and $P_{ESS,i,t}$ denote the price and power output of ESSs owned by the DSO, $C_{v,t}^{ADN}$ represents the penalty of the DSO for poor voltage quality, $\hat{V}_{i,t}$ is the voltage at the $i$th node in the ADN, $\hat{V}_{PCC,t}$ is the voltage at the Point of Common Coupling (PCC), while $\alpha$ and $\beta$ are importance coefficients determined by the DSO. For the constraints, $\Gamma_1$ denotes the voltage profile limitation, $\Gamma_2 - \Gamma_4$ denote the active power output limitation, reactive power output limitation, and ramping limitation of each CDG, respectively; $\Gamma_5$ is the power output limitation of the ESS, while $\Gamma_6$ denotes the constraint of the ESSs' State-of-charging (SOC) status; $\Gamma_7$ is the power balance constraint, indicating the total generation should be equal to the total load demand. The following equations are to compute the voltage profile at each node, as well as the total loss.

$$L_{t,loss}^{ADN} = \sum_{t \in T} \sum_{ij \in l} \frac{\rho_{ij} \left[ \left(P_{ji,t}\right)^2 + \left(Q_{ji,t}\right)^2 \right]}{\left(\hat{V}_{j,t}\right)^2} \tag{5-a}$$

$$\hat{P}_{i,t} = \sum_{w \in \delta(i)} P_{iw,t} - \sum_{j \in \varphi(j)} \left[ P_{ji,t} + \frac{\rho_{ij} \left[ \left(P_{ji,t}\right)^2 + \left(Q_{ji,t}\right)^2 \right]}{\left(\hat{V}_{j,t}\right)^2} \right] \tag{5-b}$$

$$\hat{Q}_{i,t} = \sum_{w \in \delta(i)} Q_{iw,t} - \sum_{j \in \varphi(j)} \left[ Q_{ji,t} - \frac{x_{ij} \left[ \left(P_{ji,t}\right)^2 + \left(Q_{ji,t}\right)^2 \right]}{\left(\hat{V}_{j,t}\right)^2} \right] + b_{ij} \left(\hat{V}_{i,t}\right)^2 \tag{5-c}$$

$$\left(\hat{V}_{i,t}\right)^2 = \left(\hat{V}_{j,t}\right)^2 - 2\left(\rho_{ij}P_{ji,t} + x_{ij}Q_{ji,t}\right) + \frac{\left[ \left(P_{ji,t}\right)^2 + \left(Q_{ji,t}\right)^2 \right]}{\left(\hat{V}_{j,t}\right)^2} \left(\rho_{ij}^2 + x_{ij}^2\right) \tag{5-d}$$

$$P_{iw,t} = P_{DG,i,t} + P_{ESS,i,t} + \hat{P}_{RDG,d,t} - P_{Load,i,t} \tag{5-e}$$

$$Q_{iw,t} = Q_{DG,i,t} + Q_{ESS,i,t} + \hat{Q}_{RDG,d,t} - Q_{Load,i,t} \tag{5-f}$$

where $P_{ji,t}$, $Q_{ji,t}$, $\rho_{ij}$, $x_{ij}$ and $b_{ij}$ denote the active and reactive power, resistance, reactance, and admittance in the line connecting node $i$ and $j$, respectively. $P_{iw,t}$ and $Q_{iw,t}$ are the input active and reactive power at node $i$.

**Remark.** *(Uncertain Risks in Reward Function)* Note that the common reward $\hat{R}_t$ of MG agents is a stochastic variable due to the uncertain risks faced by both MGs and the DSO:

*1) Uncertainties of MGs:* The uncertainty of affiliated RDERs $\hat{P}_{RDG,l,t}$ and load demand $\hat{P}_{MG,t}^{load}$ in each MG will result in considerable power imbalance in the MGC. Such an imbalance $\hat{P}_{MG,k,t}^{ADN}$ must be addressed by the DSO, leading to an uncertain balancing cost. If the imbalance cannot be entirely addressed, then the MGC will be penalized by $\sum_{t \in T} C_{curt,t}^{MG} \hat{P}_{curt,t}^{MG}$ due to the load curtailment.

*2) Uncertainties of DSO:* The DSO will coordinate the upstream DERs that are not affiliated to MGs. However, those RDERs are also suffering from uncertainties of power output $\hat{P}_{RDG,d,t}$, which will lead to the network imbalance of both active and reactive power. The resulted imbalance must be addressed with the support of the connected HV network with a balancing cost $\sum_{t \in T} C_{ex,t}^{HV} P_{ex,t}^{HV}$. Therefore, the charge of balancing cost $\hat{C}_{MG,t}^{ADN}$ to the MGC is also varying.

## III. ALGORITHM DESIGN AND IMPLEMENTATION

In this section, the main focus is how each MG agent optimizes its policy to maximize the common reward, while a novel RS-TRPO algorithm is proposed to tackle this problem. Firstly, the rationale of deploying TRPO as the baseline is illustrated. Then, the baseline TRPO is extended to the multi-agent environment with sequential decision-making process, being consistent with the DBC dispatch paradigm. Finally, practical implementation of this algorithm considering MGs' risk mitigation against uncertainties is elaborated.

### A. Rationale and Preliminaries

Let us firstly consider a single MG performing the self-dispatch with its own policy $\mu_t^\theta(a_t|s_t)$ parameterized by $\theta$, aiming at improving the probability of actions that result in optimal cumulative reward $\varrho(\mu_t^\theta)$ in a single day (one trajectory, 24 hours). An intuitive methodology for the optimization of $\mu_t^\theta$ is the so-called *Vanilla Policy Gradient* (VPG), which enforces $\mu_t^\theta$ to be updated via stochastic gradient ascent: $\theta_{t+1} = \theta_t + \delta \nabla_{\theta_t} \varrho(\mu_t^\theta)$, where the gradient of $\varrho(\mu_t^\theta)$ can be estimated by:

$$\varrho(\mu_t^\theta) \approx \frac{1}{T} \sum_{t \in T} \left\{ \nabla_{\theta_t} \log \mu_t^\theta(a_t|s_t) \Psi_t^\theta(s_t, a_t) \right\} \tag{6}$$

where $\Psi_t^\theta(s_t, a_t)$ is the advantage function, defined as the difference between the state-action value function (i.e., the Q-function $Q_t(s_t, a_t)$) and the state-value function $V_k(s_t)$, indicating how much better or worse the current action is compared with other actions on average.

$$\Psi_t^\theta(s_t, a_t) = \underbrace{\left\{ R_t(s_t, a_t) + \sum_{s_t \in S} \mathfrak{T}_t \Upsilon(s_t) \right\}}_{\text{① The Q-function } Q_t(s_t, a_t)} - \underbrace{\left\{ \sum_{a_t \in A} \mu_t^\theta(a_t|s_t) \left[ R_t(s_t, a_t) + \sum_{s_t \in S} \mathfrak{T}_t \Upsilon(s_{t+1}) \right] \right\}}_{\text{② The state-value function } V_k(s_t)} \tag{7}$$

However, the VPG suffers from a significant drawback that the policy is easily to trap into the local optimum or even deteriorate performance. The TRPO method [27] effectively tackles this issue by two improvements. First, the policy is updated based on the surrogate advantage, which enables the monotonically increased policy gradient in each step:

$$\theta_k^{t+1} = argmax_{\theta_k^t} \mathbb{E} \left[ \frac{\mu_k^{\theta_k^{t+1}}(a_t^k \sim \theta_k^{t+1}|s_t^k)}{\mu_k^{\theta_k^t}(a_t^k \sim \theta_k^t|s_t^k)} \Psi_k^{\theta_k^t}(s_t^k, a_t^k) \right] \tag{8}$$

Second, a KL-divergence-based constraint is imposed to limit the step size of policy update:

$$D_{KL}^k(\theta_k^t, \theta_k^{t+1}) = \mathbb{E} \left[ D_{KL}^k(\mu_k^{\theta_k^t}(\cdot|s_t^k) || \mu_k^{\theta_k^{t+1}}(\cdot|s_t^k)) \right] \tag{9}$$

By optimizing the policy within the trusted neighborhood of the current policy constrained by $D_{KL}^k(\theta_k^t, \theta_k^{t+1})$, excessively aggressive policy updates in risky directions can be avoided.

To extend the aforementioned TRPO algorithm to the cooperative dispatch of multiple MGs, one naïve approach is to construct a "joint policy" $\mu_*^\theta(a_t^1, a_t^2, ..., a_t^p | s_t^1, s_t^2, ..., s_t^p)$ among all MG agents via parameter sharing [27]. This enables each MG agent to follow the shared joint dispatch strategy parameterized by $\theta$ (but with literally different dispatch actions) based on their own observation $s_t^k$ such as total/local demands, available capacities, etc. However, this approach is trivial and impractical due to the following deficiencies. First, it requires the same action space for each individual MG agent, which is unachievable given that each MG has a unique available capacity. Second, the simultaneous dispatch of all MG agents may result in serious violation of constraints (5) due to the confliction of power flow. Third, the uncertainties that need to be tackled (as mentioned in the previous subsection) are totally neglected. These deficiencies are obstacles for the TRPO-based fully distributed dispatch paradigm to be employed in real-world MGCs, and therefore motivate us to propose a new algorithm to tackle these issues.

### B. Sequential Dispatching for Multiple MGs with Monotonic Performance Improvement

In this subsection, we extend the TRPO algorithm which guarantees the monotonic improvement of policy to the multi-MGs dispatching problem, adopting sequential decision-making paradigm with consideration of MG's own operational constraints and uncertainties. This dispatching paradigm is developed based on the following propositions that guarantee the optimality of sequential decision-making of individual MGs:

**Proposition 1.** Let $\Psi_\theta^{k_m}(s_t^k, a_t^{k_m}), m \in [1, p]$ be the joint advantage function of MG agents that reflects the performance of the joint action $a_t^{k_m}$, which represents the simultaneous dispatch orders of MGs $[1, p]$. In Markov game formulated in Table II, this advantage function is decomposed as follows:

$$\Psi_\theta^{k_m}(s_t^k, a_t^{k_{1:p}}) = \sum_{m=1}^{p} \Psi_\theta^{k_m}(s_t^k, a_t^{k_{1:p-1}}, a_t^{k_p}) > 0 \tag{10}$$

which indicates that the sequential dispatch of individual MGs following an arbitrary order could be equivalent to the simultaneous dispatch of all MG agents. Specifically, for the arbitrary order $k_{1:p}$, MG $k_1$ firstly observes state $s_t^{k_1}$ and take dispatch orders $a_t^{k_1}$, resulting in an individual advantage function $\Psi_\theta^{k_1}(s_t^{k_1}, a_t^{k_1}) > 0$. Then, MG $k_2$ makes its dispatch order $a_t^{k_2}$ based on both its observation $s_t^{k_2}$ and the former MG $k_1$'s dispatch order $a_t^{k_1}$, resulting in the advantage function $\Psi_\theta^{k_2}(s_t^{k_2}, a_t^{k_1}, a_t^{k_2}) > 0$. Similarly, MGs $k_{3,...,p}$ sequentially make their own dispatch order based on their own observations and all of the former MGs' dispatch orders with positive $\Psi_\theta^{k_m}(s_t^k, a_t^{k_{1:p-1}}, a_t^{k_p})$. We will then extend this proposition to the joint *dispatch strategy* rather than joint *dispatch actions*.

**Proposition 2.** Let $\hat{\mu}_{\theta_m}^{k_{1:m}} = \{\hat{\mu}_{\theta_1}^{k_1}, \hat{\mu}_{\theta_2}^{k_2}, ..., \hat{\mu}_{\theta_m}^{k_m}\}, \forall m \in [2, p]$ denote the sequentially-updated policy of MGs $k_{1:p}$, based on the previous policy $\mu_{\theta_m}^{k_m}$. For the performance of $\hat{\mu}_{\theta_m}^{k_{1:m}}$ in comparison with $\hat{\mu}_{\theta_{m-1}}^{k_{1:m-1}}$, i.e., at each step of sequential update, the inequality $\varrho(\hat{\mu}_{\theta_m}^{k_{1:m}}) \geq \varrho(\hat{\mu}_{\theta_{m-1}}^{k_{1:m-1}})$ always holds.

**Proof.** Consider the group of MG agents $k_{1:m-1}$ who already have sequentially updated their dispatch strategies $\hat{\mu}_{\theta_{m-1}}^{k_{1:m-1}} = \{\hat{\mu}_{\theta_1}^{k_1}, \hat{\mu}_{\theta_2}^{k_2}, ..., \hat{\mu}_{\theta_{m-1}}^{k_{m-1}}\}$, while the MG $k_p$ is to randomly update its dispatch strategy $\tilde{\mu}_{\theta_m}^{k_m}$. Define the *surrogate advantage* of such a dispatch strategy update:

$$\mho_{\theta_m}^{k_{1:m}}(\hat{\mu}_{\theta_{m-1}}^{k_{1:m-1}}, \tilde{\mu}_{\theta_m}^{k_m})$$
$$= \mathbb{E}\left[\Psi_\theta^{k_m}(s_t^{k_m}, \hat{a}_t^{k_{1:m-1}} \sim \hat{\mu}_{\theta_{m-1}}^{k_{1:m-1}}, \tilde{a}_t^{k_m} \sim \tilde{\mu}_{\theta_m}^{k_m})\right], \forall m \in [2, p] \tag{11}$$

Based on Theorem 1 of [27], it can be obtained that:

$$\varrho\left(\hat{\mu}_{\theta_m}^{k_m}\right) \geq \varrho\left(\hat{\mu}_{\theta_{m-1}}^{k_{1:m-1}}\right) + \mho_{\theta_m}^{k_{1:m}}\left(\hat{\mu}_{\theta_{m-1}}^{k_{1:m-1}}, \tilde{\mu}_{\theta_m}^{k_m}\right) \tag{12}$$

in which $\mho_{\theta_m}^{k_{1:m}}\left(\hat{\mu}_{\theta_{m-1}}^{k_{1:m-1}}, \tilde{\mu}_{\theta_m}^{k_m}\right) \geq 0$ (see Proposition 1) while the equality holds if $\hat{\pi}_{\theta_{m-1}}^{k_{1:m-1}} = \tilde{\pi}_{\theta_m}^{k_m}$, i.e., the $m$th MG agent makes no update of its dispatch strategy. This concludes the proof. Intuitively, the monotonic improvement of $\varrho\left(\hat{\mu}_{\theta_m}^{k_{1:m}}\right)$ at each step $m \in [2, p]$ results in a significant cumulative superiority when the last MG agent finishes its policy update:

$$\varrho\left(\hat{\mu}_{\theta_m}^{k_m}\right) - \varrho\left(\mu_{\theta_m}^{k_m}\right) \geq \sum_{m=2}^{p} \mho_{\theta_m}^{k_{1:m}}\left(\hat{\mu}_{\theta_{1:m-1}}^{k_{1:m-1}}, \tilde{\mu}_{\theta_m}^{k_m}\right) \tag{13}$$

where $\varrho\left(\mu_{\theta_m}^{k_m}\right) = \mathbb{E}\left[\Psi_\mu^{k_m}(s_t^{k_m}, a_t^{k_m} \sim \mu_{\theta_m}^{k_m})\right]$.

Based on these propositions, we can now extend the single agent TRPO to the multi-MGs dispatching, by enabling each MG agent to update its dispatch strategy sequentially:

$$\hat{\mu}_{\theta_m}^{k_m} = argmax_{\hat{\mu}_{\theta_m}^{k_m}} \mho_{\theta_m}^{k_{1:m}}\left(\hat{\mu}_{\theta_{1:m-1}}^{k_{1:m-1}}, \tilde{\mu}_{\theta_m}^{k_m}\right)$$
$$= argmax_{\hat{\mu}_{\theta_m}^{k_m}} \mathbb{E}\left[\Psi_\mu^{k_m}(s_t^{k_m}, a_{\theta_{1:m}}^{k_{1:m-1}} \sim \hat{\mu}_{\theta_{m-1}}^{k_{1:m-1}}, a_t^{k_m} \sim \tilde{\mu}_{\theta_m}^{k_m})\right] \tag{14}$$

with KL constraint $D_{KL}^m(\theta_{k_m}^{new}, \tilde{\theta}_{k_m}^{new}) \leq \varepsilon$ to limit step size. To be more concrete, the $m$th MG agent updates its dispatch strategy $\tilde{\mu}_{\theta_m}^{k_m}$ with the observation of both current state $s_t^{k_m}$ and previous $m-1$ MG agents' updated strategy $\hat{\mu}_{\theta_{1:m-1}}^{k_{1:m-1}}$, in order to maximize the advantage $\Psi_\mu^{k_m}$ parameterized by the joint policy $\mu$ at the previous iteration.

### C. Practical Implementation with Risk Mitigation

Based on the aforementioned theoretical foundation of the sequential dispatching under the framework of TRPO, the practical algorithm design is proposed in this section to answer two main questions: 1) How to design a practical algorithm for MGs to implement the sequential dispatch, as formulated in (14)? 2) How to deal with the risk of MGs, as discussed in Section Ⅱ. B? Briefly, the *actor-critic* network is adopted herein to implement MGs' individual policy gradient, and the CVaR is incorporated in the policy gradient (both the actor and critic network) to facilitate the risk mitigation.

*1) Actor Network*: This network is designated to update MGs' policy $\tilde{\mu}_{\theta_m}^{k_m}$ based on their "performance" $\Psi_\mu^{k_m}$ estimated by critic network. However, the gradient $\nabla_{\tilde{\mu}_{\theta_m}^{k_m}} \mathbb{E}\left[\Psi_\mu^{k_m}(\hat{\mu}_{\theta_{1:m-1}}^{k_{1:m-1}})\right]$ is very hard to be computed since it involves a function composition with three policies: $\mu$, $\hat{\mu}$ and $\tilde{\mu}$. Here, the *importance sampling* [27] and the *chain rule* is adopted to tackle this problem:

$$\nabla_{\theta_m} \mathbb{E}\left[\Psi_{\boldsymbol{\mu}}^{k_m}(\hat{\boldsymbol{\mu}}_{\theta_{1:m-1}}^{k_{1:m-1}})\right] = \nabla_{\boldsymbol{\mu}} \Psi_{\boldsymbol{\mu}}^{k_m} \cdot \mathbb{E}\left[\Psi_{\boldsymbol{\mu}}^{k_m} \cdot \Lambda_{\boldsymbol{\mu}}^{k_m}\right] \quad (15)$$

where

$$\Lambda_{\boldsymbol{\mu}}^{k_m} = \frac{\nabla_{\theta_m} \tilde{\mu}_{\theta_m}^{k_m}\left(a_{\theta_m}^{k_m}\Big|s_t^{k_m}\right)}{\mu_{\theta_m}^{k_m}\left(a_{\theta_m}^{k_m}\Big|s_t^{k_m}\right)} \cdot \frac{\hat{\boldsymbol{\mu}}_{\theta_{1:m-1}}^{k_{1:m-1}}\left(a_{\theta_{1:m-1}}^{k_{1:m-1}}\Big|s_t^{k_m}\right)}{\boldsymbol{\mu}_{\theta_{1:m-1}}^{k_{1:m-1}}\left(a_{\theta_{1:m-1}}^{k_{1:m-1}}\Big|s_t^{k_m}\right)}$$
$$= \hat{\Lambda}_{\boldsymbol{\mu}}^{k_m} \cdot \nabla_{\theta_m} log \tilde{\mu}_{\theta_m}^{k_m}\left(a_{\theta_m}^{k_m}\Big|s_t^{k_m}\right) \cdot \Psi_{\boldsymbol{\mu}}^{k_m} \quad (16)$$

in which

$$\hat{\Lambda}_{\boldsymbol{\mu}}^{k_m} = \frac{\tilde{\mu}_{\theta_m}^{k_m}\left(a_{\theta_m}^{k_m}\Big|s_t^{k_m}\right)}{\mu_{\theta_m}^{k_m}\left(a_{\theta_m}^{k_m}\Big|s_t^{k_m}\right)} \cdot \frac{\hat{\boldsymbol{\mu}}_{\theta_{1:m-1}}^{k_{1:m-1}}\left(a_{\theta_{1:m-1}}^{k_{1:m-1}}\Big|s_t^{k_m}\right)}{\boldsymbol{\mu}_{\theta_{1:m-1}}^{k_{1:m-1}}\left(a_{\theta_{1:m-1}}^{k_{1:m-1}}\Big|s_t^{k_m}\right)} \quad (17)$$

In (16), the input tuples $\{s_t^{k_m}, a_{\theta_{1:m-1}}^{k_{1:m-1}}, a_{\theta_m}^{k_m}\}$ are sampled from MGs' previous policy $\boldsymbol{\mu}$. Thus, it avoids computing $\hat{\boldsymbol{\mu}}_{\theta_{1:m-1}}^{k_{1:m-1}}$ and $\tilde{\mu}_{\theta_m}^{k_m}$ based on $a_{\theta_{1:m-1}}^{k_{1:m-1}} \sim \hat{\boldsymbol{\mu}}$ with overwhelming computational burden, and $a_{\theta_m}^{k_m} \sim \tilde{\mu}$ which is unknown.

However, computing the term $\nabla_{\theta_m} \Psi_{\boldsymbol{\mu}}^{k_m}$ is still a challenge, because MG's common reward function $R_t$ is stochastic with uncertainty set $\Omega$, as discussed in Section II. B, and therefore $\Psi_{\boldsymbol{\mu}}^{k_m}(R_t)$ is also a random variable with the cumulative density function (CDF): $\mathbb{C}[\Psi_{\boldsymbol{\mu},X}^{k_m}(x)] = \Pr(X \leq x)$. Intuitively, the "risk" faced by the MGC refers to $X$ being lower a threshold value, indicating an unacceptable performance of a specified policy. Such a risk can be defined using CVaR as follows:

**Definition 3.** *(CVaR-based Advantage Function)* Given the $\alpha$ confidence level, the $\alpha$-CVaR of $X \sim \Psi_{\boldsymbol{\mu},X}^{k_m}(x)$ is defined as the expectation of the worst $\alpha \in [0,1]$ quantile of $X$:

$$\Theta_\alpha(\Psi_{\boldsymbol{\mu},X}^{k_m}) := \mathbb{E}[X|X \leq \inf\{x : \mathbb{C}[\Psi_{\boldsymbol{\mu},X}^{k_m}(x)] \geq \alpha\}] \quad (18)$$

where $\inf\{x : \mathbb{C}[\Psi_{\boldsymbol{\mu},X}^{k_m}(x)] \geq \alpha\}$ denotes the $\alpha$-quantile VaR of $X$, denoting the "threshold value" that can be tolerated. The parameter $\alpha$ indicates the level of tolerance to the risk; when $\alpha = 1$, (18) actually equivalents to the original $\Psi_{\boldsymbol{\mu},X}^{k_m}$ without any consideration of risk.

**Proposition 3.** *(Policy Gradient of $\alpha$-CVaR)* With the above definition, the gradient of $\Psi_{\boldsymbol{\mu}}^{k_m}$ w.r.t. $\theta_m$ with consideration of CVaR as a risk-sensitive manner can be computed by:

$$\nabla_{\theta_m} \Psi_{\boldsymbol{\mu}}^{k_m} \approx \frac{1}{\alpha DT} \sum_{\alpha N} \sum_{t \in T} \nabla_{\theta_m} log \tilde{\mu}_{\theta_m}^{k_m}\left(a_{\theta_m}^{k_m}\Big|s_t^{k_m}\right) \cdot (\Psi_{\boldsymbol{\mu}}^{k_m} - \mathbb{C}) \quad (19)$$

where $\alpha N$ indicates the worst $\alpha$-quantile samples among all the trajectories with a batch size of $D$, and $\mathbb{C}$ denotes the estimation of $\alpha$-VaR. For proof please see [28].

Similar to the single-agent TRPO, we further simplify the objective of gradient descent in (19) using Taylor expansion:

$$\tilde{\theta}_{k_m}^{new} = \theta_{k_m}^{old} + \sqrt{\frac{2\varepsilon\rho^2 j}{\nabla_{\tilde{\mu}} \cdot (H_{k_m})^{-1} \cdot \nabla_{\tilde{\mu}}}} \cdot (H_{k_m})^{-1} \cdot \nabla_{\tilde{\mu}} \quad (20)$$

where $H_{k_m}$ denotes the Hessian of $\nabla_{\theta_{k_m}^{old}}^2 D_{KL}^m\left(\theta_{k_m}^{old}, \tilde{\theta}_{k_m}^{new}\right)$, and $\nabla_{\tilde{\mu}}$ is a simplified notation of the gradient (19).

*2) Critic Network:* This network is designated to update the estimation of $\mathbb{E}[\Psi_{\boldsymbol{\mu}}^{k_m}]$ as the performance of MGs' policy, to facilitate computing the gradient in (19). Note that $\Psi_{\boldsymbol{\mu}}^{k_m}$ is a joint advantage function, which means that we only need to maintain one centralized critic network instead of distributed critic networks for each MG agent. In this algorithm, the Temporal-Difference (TD) error, which has the same expectation as the advantage function [27], is applied to estimate the joint advantage function $\Psi_{\boldsymbol{\mu}}^{k_m}$:

$$\Psi_{\boldsymbol{\mu}}^{k_m} \approx R_t + \gamma V_\varphi(s_{t+1}) - V_\varphi(s_t) \quad (21)$$

where $V_\varphi$ is estimated by the critic parameterized by $\varphi$:

$$\varphi_{t+1} = argmin_\varphi \frac{1}{DT} \sum_{d \in D} \sum_{t \in T} [V_\varphi(s_t) - R_t]^2 \quad (22)$$

The pseudo code of implementing this algorithm is summarized in Algorithm 1.

---

**Algorithm 1** Workflow of RS-TRPO Algorithm

**Input:**
    1) Network configuration parameters and line capacity limitations;
    2) Rated capacity and ramping limitation of generators;
    3) Hyper parameters as listed in Table III.

**Output:** Dispatch policies of each MG.

**for** $t = 0,1,2,\ldots,T-1$ **do**

1: Generate the trajectories $\{s_t^k, a_t^{k_m}, r_t^k, s_{t+1}^k\}$ based on the inputted load demand profile, the uncertainty set, and the joint policy $\boldsymbol{\mu}_t^k$.

2: Sample a mini-batch of transitions from replay buffer $D$.

3: Compute the advantage function based on (18).

4: Generate the random sequence for MGs' dispatch order.

    **for** agent $k = 1,2,\ldots,p$ **do**

5: Compute the gradient of agent's policy using (19).

6: Update agent's policy (i.e., actor network) using (20).

    **end for**

7: Update the critic network using (21) – (22).

**end for**

---

## IV. CASE STUDY

### A. Test System and Input Data

In this case study, a test MGC consisting of 4 autonomous MG agents in a modified IEEE 33-Bus Network is adopted as shown in Fig.5, with the following input data to be generated as indicated in Algorithm 1. The rated capacity and generation cost of different generators are referred to [29]. The prediction error of power output of wind turbines (WTs), photovoltaics (PVs) and the system load demand is generated based on the probability density function (PDF) provided in [30] with consideration of PDF correlations among DERs and the load demand. The following three baseline algorithms will also be performed and compared with the proposed RS-TRPO algorithm, including the MATRPO [27] and MADDPG [31] as representatives of MARL without guarantee of optimality and risk mitigation, and to formulate the model as the equilibrium problem with equilibrium constraints (EPEC) [32], as a representative of non-learning-based methods. Detailed instructions of the implementation of these algorithms can be found in [31] – [32]. Note that in EPEC, we model the MGs' interactions as a competitive game, in which each MG's utility function is composed by the MGC's total profit plus MG's own penalties. Key hyper parameters for implementing these algorithms are summarized in Table III.

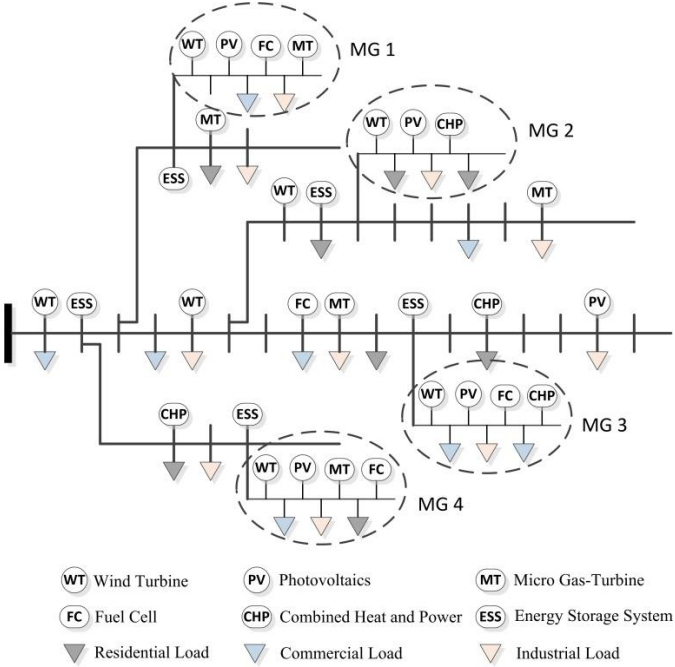

WT Wind Turbine    PV Photovoltaics    MT Micro Gas-Turbine

FC Fuel Cell    CHP Combined Heat and Power    ESS Energy Storage System

▼ Residential Load    ▼ Commercial Load    ▼ Industrial Load

Fig.2. Modified IEEE 33-Node Test System

TABLE III
HYPER PARAMETERS REGARDING TO ALGORITHM IMPLEMENTATIONS

| Algorithm | Key hyper parameters |
|---|---|
| *RS-TRPO* | • Network dimensions and layers: 2 layers, with dimension of 64 and 32<br>• Batch size: 64<br>• Capacity of experience replay buffer: 1e8<br>• Learning rate: 1e-4<br>• Discount factor: 0.9 |
| *MATRPO and MADDPG* | • In order to keep the fairness, all the hyper parameters are same as those in the RS-TRPO algorithm. |

### B. Computational Performance and Sensitivity Analysis in Terms of Penalties for Constraint Violations

In this section, the computational performance of different algorithms is compared in terms of the reward and voltage profile. Fig.3 and Fig.4 demonstrates the reward of these 4 algorithms in 4 scenarios, and Table IV provides the average reward of different MG agents, while scenarios 1-4 denote: 1) with low load demand (50% of the rated capacity) and low uncertainty (50% of the peak value of total prediction error in [30]); 2) with low load demand and high uncertainty; 3) with high load demand and low uncertainty; and 4) with high load demand and high uncertainty.

We firstly assume a high penalty for active power balancing (500RMB/MW) and a low penalty (100RMB/MW) for voltage violation (i.e., nodal voltage higher than 1.05 p.u. or lower than 0.95 p.u. after MGs' dispatch). The average rewards for each algorithm across all four scenarios are as follows: RS-TRPO (-

140.65 RMB), MADDPG (-170.46 RMB), MATRPO (-162.91 RMB), and EPEC (-155.30 RMB). On average, RS-TRPO outperforms MADDPG by 17.6%, MATRPO by 13.75%, and EPEC by 9.55%. This consistent superior performance suggests that RS-TRPO is an effective algorithm for optimizing the dispatch strategy of MGs. The decentralized learning and risk mitigation capability of our proposed algorithm likely plays a significant role in its superior performance, enabling it to manage the distributed dispatch and net load uncertainties better than the other algorithms considered in this comparison.

On average across all four scenarios, MADDPG (-170.46 RMB) yields the lowest average rewards, underperforming MATRPO by 17.59%, MATRPO by 9.01%, and EPEC by 8.69%. This could be attributed to its lack of risk mitigation, and the lack of guarantee of monotonically increased policy gradient (as discussed in Section III. A). MATRPO (-162.91 RMB) performs better than MADDPG but still falls behind RS-TRPO and EPEC, underperforming them by 13.75% and 3.68% on average. EPEC (-155.30 RMB) shows relatively better performance in comparison with MADDPG and MATRPO, outperforming MADDPG by 8.69% and MATRPO by 3.68% on average. Its formulation enables it to consider interactions between multiple microgrids. However, EPEC's approach to solving the problem by finding an equilibrium point in the decision-making process, while taking into account the constraints and interactions of multiple microgrids, might be less adaptive to uncertainties in the net load than the learning-based methods like RS-TRPO. As a result, EPEC underperforms MATRPO by 9.55% on average.

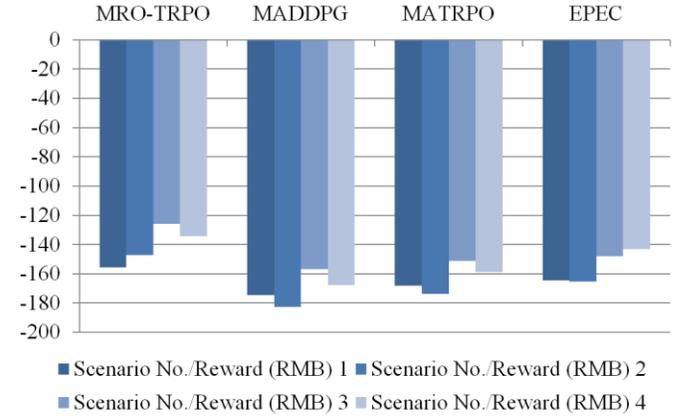

■ Scenario No./Reward (RMB) 1    ■ Scenario No./Reward (RMB) 2
■ Scenario No./Reward (RMB) 3    ■ Scenario No./Reward (RMB) 4

Fig.3. Average Reward of Different Algorithms (with a high penalty for active power balancing and a low penalty for voltage violation)

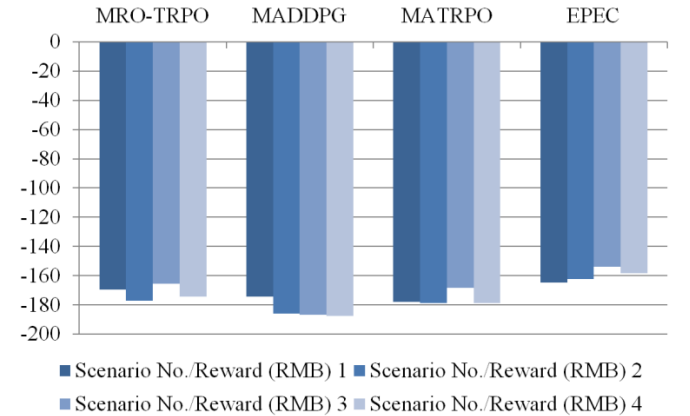

■ Scenario No./Reward (RMB) 1    ■ Scenario No./Reward (RMB) 2
■ Scenario No./Reward (RMB) 3    ■ Scenario No./Reward (RMB) 4

Fig.4. Average Reward of Different Algorithms (with a low penalty for active power balancing and a high penalty for voltage violation)

TABLE IV
AVERAGE VOLTAGE VIOLATION OF DIFFERENT ALGORITHMS

| Algorithm | Scenario No./Voltage (p.u.) | | | |
|---|---|---|---|---|
| | 1 | 2 | 3 | 4 |
| MADDPG | 0.023 | 0.031 | 0.018 | 0.029 |
| RS-TRPO | 0.015 | 0.019 | 0.023 | 0.022 |
| MATRPO | 0.019 | 0.028 | 0.024 | 0.027 |
| EPEC | 0.003 | 0.012 | 0.004 | 0.010 |

However, when analyzing the performance of these methods under a low penalty for active power balancing (100RMB/MW) and a high penalty (1000RMB/p.u.) for voltage violation, the proposed RS-TRPO algorithm shows a worse performance in terms of total reward than EPEC. This is due to the fact that EPEC has a better guarantee of voltage constraints satisfaction than RL methods (shown in Table IV), as the relationship between actions in RL (i.e., power output) and nodal voltages is implicit and thus cannot be incorporated in RL. This is a significant deficiency to be addressed in our future work by adopting Safe RL.

### C. Impact of Risk Preferences on Decision Making

This subsection explores the impact of low and high risk preferences on the generation output of different microgrid (MG) agents using the proposed RS-TRPO algorithm and the EPEC method which does not incorporate the capability of risk mitigation.

By using RS-TRPO, under low risk preferences ($\alpha = 0.9$), MG agents tend to allocate a larger share of their generation output to renewable sources, such as WT and PV, as shown in Fig.6. This is exemplified in Scenario 1 (0.712 out of 1.314, 53.85% renewables), Scenario 2 (0.736 out of 1.725, 41.18% renewables), Scenario 3 (1.421 out of 1.606, 87.50% renewables), and Scenario 4 (1.403 out of 2.112, 66.67% renewables). In these scenarios, MG agents capitalize on the cost-saving potential of renewables while acknowledging the inherent variability in their output. In contrast, high risk preferences ($\alpha = 0.5$) prompt MG agents to reduce reliance on renewable generation in order to minimize additional balancing costs, as shown in Fig.7. This behavior is demonstrated in Scenario 1 (0.302 out of 1.204, 25% renewables), Scenario 2 (0.315 out of 1.408, 21.43% renewables), Scenario 3 (0.621 out of 1.023, 62.15% renewables), and Scenario 4 (0.617 out of 1.211, 50.23% renewables). By decreasing the proportion of renewable sources, MG agents mitigate the potential imbalance costs associated with fluctuating generation output.

Fig.8 displays the generation output of MG agents using the EPEC. When comparing EPEC with RS-TRPO's high risk preference, the two algorithms demonstrate a similar aggressive approach to dispatching renewable energy sources, such as wind turbines (WT) and photovoltaic (PV) systems. For example, in Scenario 1, the EPEC algorithm allocates 0.4 WT and 0.3 PV for MG Agent 1 (58.33% of the total generation output) to renewables, while RS-TRPO's high risk preference assigns 0 WT and 0.3 PV (25% of the total generation output). In Scenario 2, EPEC's renewable share is 0.4 WT and 0.3 PV for MG Agent 1 (50% of the generation output), closely resembling the 0.3 PV (21.43%) share in RS-TRPO's high risk preference.

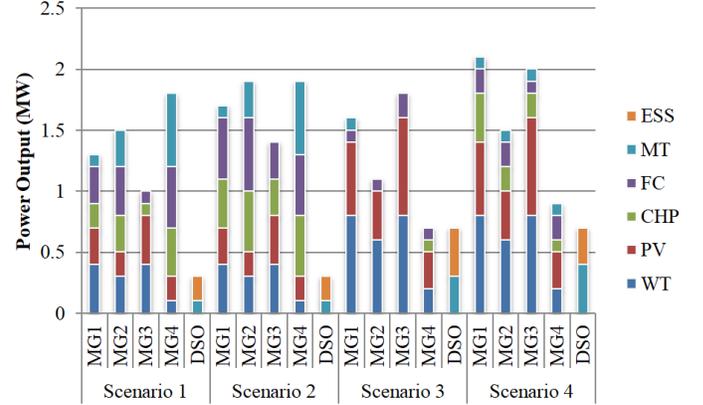

Fig.5. Power Output by Implementing the RS-TRPO Algorithm with a High Risk Preference

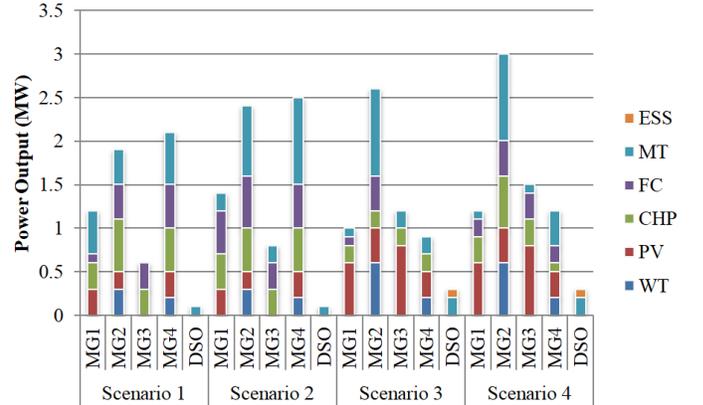

Fig.6. Power Output by Implementing the RS-TRPO Algorithm with a Low Risk Preference

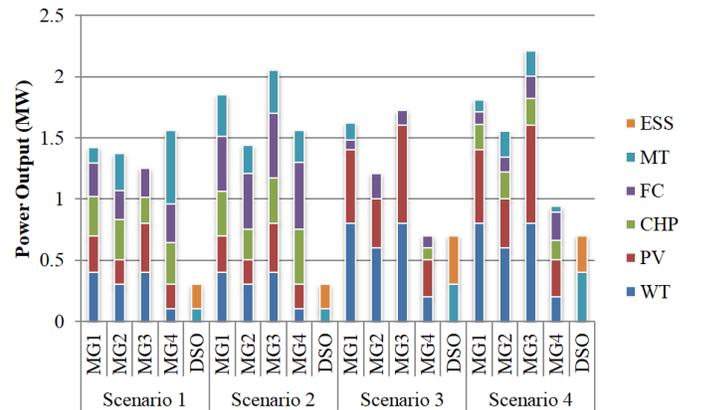

Fig.7. Power Output by Implementing the EPEC Method

In contrast, MATRPO's low risk preference demonstrates a more conservative approach when compared to EPEC. For instance, in Scenario 1, the renewable energy share for RS-TRPO's low risk preference for MG Agent 1 is 0.4 WT and 0.3 PV (53.85% of the generation output), whereas EPEC assigns 0.4 WT and 0.3 PV (57.69% of the generation output) to renewables. Similarly, in Scenario 4, RS-TRPO's low risk preference allocates 0.8 WT and 0.6 PV for MG Agent 1 (50% of the generation output) to renewables, while EPEC has a higher share of 0.8 WT and 0.6 PV (66.67% of the generation output). This indicates that RS-TRPO's low risk preference opts for a more cautious strategy in deploying renewable to reduce uncertainties and additional balancing costs.

## V. Conclusion

The integration of MGs into MGCs enhances the reliability and flexibility of energy supply through resource sharing, but the dispatch of MGCs presents a challenge for their secure and economic operation. In this paper, a novel RS-TRPO has been proposed to tackle this problem, enabling autonomous MGs in the MGC to implement their self-dispatch and mitigate potential conflicts. The algorithm considers multiple objectives (cost and voltage profile) and risk mitigation. The results show that the proposed approach can effectively optimize the operation of the MGC while reducing operating costs and improving system reliability. Specifically, the proposed approach outperforms other optimization methods, including MADDPG, PSO and EPEC, in terms of economic efficiency, voltage profile and risk mitigation. However, the proposed method has the drawback to deal with implicit constraints such as the voltage limitation, which should be tackled in future research. Overall, these findings suggest that the proposed approach has significant potential for optimizing MGCs and improving their economic benefits.